# Probing Dy$^{3+}$ magnetic moments in multiferroic perovskite DyMnO$_3$ by optical spectroscopy

**M.A. Kashchenko**[1,2], **S.A. Klimin**[1], **A. M. Balbashov**[3], **and M.N. Popova**[1]

[1] Institute of Spectroscopy, Russian Academy of Sciences, 5, Fizicheskaya Str. 108840 Troitsk, Moscow, Russia
[2] Moscow Institute of Physics and Technology (State University), 141700 Dolgoprudnyi, Moscow region, Russia
[3] Moscow Power Engineering Institute (Technical University), 14, Krasnokazarmennaya Str., 111250 Moscow, Russia

We present a detailed temperature-dependent (4 – 300 K) spectroscopic study of DyMnO$_3$ single crystals with distorted perovskite structure. Energies of 36 crystal-field levels of Dy$^{3+}$ in paramagnetic DyMnO$_3$ were determined. The Dy$^{3+}$ ground Kramers doublet does not split at $T_N^{Mn}$ = 39 K and splits below $T_{lock}$=18 K. The splitting grows fast at temperatures near $T_N^{Dy}$ =6.5 K and reaches $\Delta_0 \approx 11\pm 2$ cm$^{-1}$ at 4 K. Using the experimental temperature dependence $\Delta_0(T)$, we calculate the dysprosium magnetic moment $m_{Dy}(T)$ and the dysprosium contribution into specific heat and magnetic susceptibility. Analysing all the experimental data, we conclude that the Dy-Mn interaction is of the Dzyaloshinskii-Moriya type.

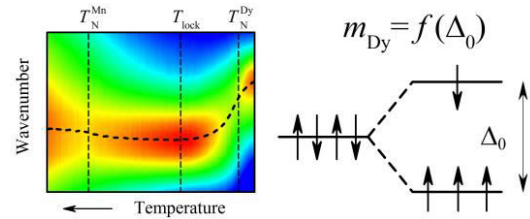

Intensity map in the temperature - wave number coordinates for a spectral line corresponding to the *f-f* transition of Dy$^{3+}$ in DyMnO$_3$ and a scheme of the splitting of the Dy ground Kramers doublet. Arrows represent Dy magnetic moments.

Copyright line will be provided by the publisher

**1 Introduction** Orthorhombic DyMnO$_3$ (as well as TbMnO$_3$) is a multiferroic, in which the ferroelectric polarization is induced by a cycloidal magnetic order through the inverse Dzyaloshinskii-Moriya (DM) interaction [1-4]. DyMnO$_3$ has an orthorhombically distorted perovskite structure and belongs to the *Pbnm* space group ($Z$=4, $a$=5.278, $b$=5.834, $c$=7.378 Å [5]). The Mn$^{3+}$ ions reside in a single fourfold *4b* $C_i$ symmetry octahedral position. The Dy$^{3+}$ ions occupy *4c* positions having the $C_s$ symmetry [5]. The Mn$^{3+}$ magnetic moments order at $T_N^{Mn}$ = 39 K along the *b* axis in a sinusoidal incommensurate (ICM) antiferromagnetic (AFM) structure with the propagation vector (0 $k_{Mn}$ 0) [6-9]. A ferroelectric polarization was not detected in this phase [6,8]. It has been shown recently [10] that the paraelectric sinusoidal phase in DyMnO$_3$ should be interpreted as a dynamical equilibrium of magnetic cycloids with opposite chiralities. These cycloids generate oppositely directed ferroelectric polarizations ±**P**$_c$ along the *c* axis, which simulates a paraelectric state. At $T_{lock}$=18 K, the magnetic structure locks into a single *bc* cycloid [9] and a ferroelectric order establishes with **P**||*c* [6,8]. A coupling between the Dy and Mn magnetic moments induces a magnetic ordering of Dy below $T_{lock}$ [8]. In their turn, Dy magnetic moments increase the amplitude of the Mn spin spiral and in such a way enhance the ferroelectric polarization of DyMnO$_3$ [8].

Below $T_N^{Dy}$ = 6.5 K, the Dy magnetic moments order in a commensurate (CM) structure with the propagation vector (0 ½ 0), different from that of Mn [8,9]. At this temperature, the ferroelectric polarization sharply drops [6,8]. It can be restored by applying a magnetic field ~ 2 T along the *a* axis [6,11] or by adding of about 20% of holmium [12]. In both cases, an independent Dy collinear magnetic ordering is suppressed.

In recent optical [13] and neutron scattering [9] experiments magnetic excitations of DyMnO$_3$ were studied. Two branches were found (with energies ~ 2 and 3 meV at *k*=0) and attributed to electromagnons. It is of interest to explore a possible interaction of electromagnons with crystal-field (CF) excitations. Information on CF levels and wave functions (which can be obtained from the CF calculations) of Dy in DyMnO$_3$ is essential also for a deeper understanding of the mechanism of multiferroicity in this compound.

As Dy$^{3+}$ is a Kramers ion and resides in the position with symmetry lower than cubic, free-ion Dy$^{3+}$ levels with a given total momentum *J* split into *J* +1/2 CF Kramers doublets, the degeneracy of which can be lifted only by magnetic interactions. Little is known on Dy CF levels and their splitting in magnetically ordered states of DyMnO$_3$. Infrared transmission study of DyMnO$_3$ at several temperatures [14] reports on 5 out of 7 CF levels of the first excit-ed $^6H_{13/2}$ multiplet and gives the values 30 and 185 cm$^{-1}$ for the splitting of the ground Kramers doublet below $T_N^{Mn} = 39$ K and the energy of the next Kramers doublet at 100 K, respectively. However, these two latter values cannot be determined unambiguously and reliably without studying a detailed temperature dependence of spectral lines. For CF calculations, it is also desirable to have as many experimental CF energies as possible.

In the present work, we perform a detailed temperature-dependent spectroscopic study of DyMnO$_3$ single crystals, with the aim to find CF energies and to probe Dy$^{3+}$ magnetic moments. The magnetic moment of a Kramers ion (like Dy$^{3+}$) is intrinsically related to the splitting of its ground doublet $\Delta_0$ and is proportional to the difference of populations of the split components.

**2 Experimental** The crystals were grown in air by floating-zone technique on the crystal-growth equipment URN-2-ZM [15], using seeds oriented along the *c* axis. The crystallization speed was about 10 mm/h, crystals were annealed at 1200° C in the growth process. Thus grown single crystals were 6-7 mm in diameter and 60-70 mm in length. Quality of the crystals was inspected by x-ray topography and Laue survey. A pair of thin single-crystalline plates of o-DyMnO$_3$ cut parallel to the (*ac*) and (*bc*) planes, with approximate size 2×2×0.1 mm was used for spectroscopic measurements in polarized light. To get information on dysprosium multiplets in the region of a strong absorption of manganese, KBr pellet with a small content of o-DyMnO$_3$ (effective thickness 10 mkm) was prepared. A Bruker IFS125HR Fourier spectrometer was used to measure the spectra in a wide region from 2000 to 10000 cm$^{-1}$. The spectral resolution up to 0.5 cm$^{-1}$ was chosen to reproduce correctly all details in the spectra. For low-temperature studies, the samples were put into a Cryomech PT403 helium closed-cycle optical cryostat.

**3 Results and discussion** Figure 1a displays a broadband transmission spectrum of DyMnO$_3$ at two temperatures, $48 > T_N^{Mn}$ and 5 K. Two groups of sharp absorption lines due to *f-f* transitions of Dy$^{3+}$, centered at 3700 and 6050 cm$^{-1}$, are superimposed onto a wing of a broad strong absorption band related to manganese. The most pronounced temperature changes of this wing occur at ~ 5500 cm$^{-1}$, so we have plotted the transmittance at this wavenumber as function of temperature (Inset of Fig. 1a). Two peculiarities, at the temperatures $T_N^{Mn} = 39$ K and $T_N^{Dy} = 6.5$ K, are clearly visible, which testifies changes of the manganese energy spectrum at an ordering of both Mn$^{3+}$ and Dy$^{3+}$ magnetic subsystems and gives a clear evidence of interaction between these two subsystems.

Using KBr pellet with small content of DyMnO$_3$, *f-f* transitions up to 10000 cm$^{-1}$ were registered. Figs. 1b-d present detailed spectra of several Dy$^{3+}$ multiplets, obtained by subtracting a broad background due to Mn$^{3+}$ absorption. Table 1 lists the energies of 36 CF levels extracted from the spectra of a paramagnetic phase. In Ref. [14], the only CF levels presented were one and five ones for the $^6H_{15/2}$ and $^6H_{13/2}$ multiplets, respectively. Our data for $^6H_{13/2}$ are consistent with those of Ref. [14], whereas we correct the position of the lowest excited level of the ground multiplet and instead of one CF level near 185 cm$^{-1}$ resolve two close levels. Several lowest excited CF levels of the ground multiplet were found from the temperature-dependent spectra (the lines starting from excited levels freeze out with lowering the temperature, see, e.g., Fig. 1b). Reliability of the first excited level at 107 cm$^{-1}$ was confirmed by a comparison (Inset of Fig. 1b) of the integral intensity $I(T)$ calculated from the absorption coefficient for a line corresponding to a transition starting at this level with energy $E_i$ and the population $n_i(T)$ of the level, calculated according to the Boltzman distribution among the set of levels of Table 1:

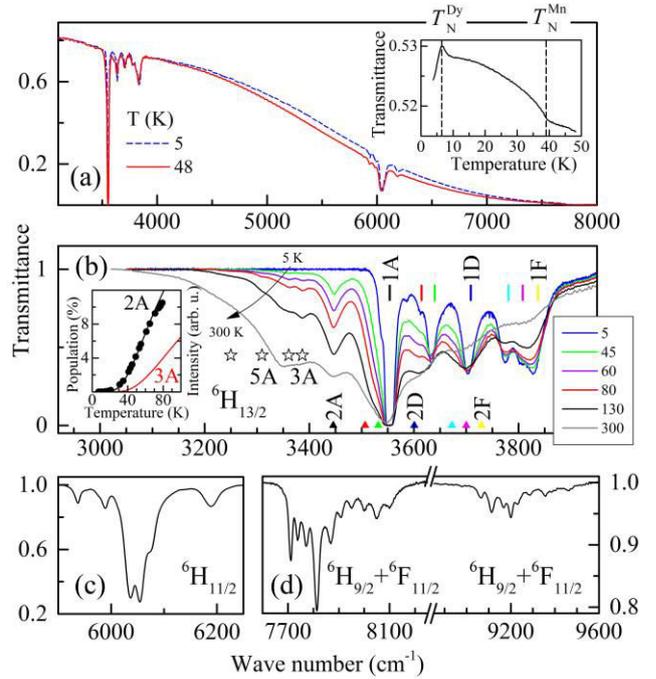

**Figure 1** Transmission spectra of DyMnO$_3$ (a) in a broad frequency range in a paramagnetic paraelectric state (48 K) and in magnetically ordered ferroelectric state at 5 K. Narrow lines are due to *f-f* transitions of Dy$^{3+}$. Inset: the temperature dependence of transmittance at 5500 cm$^{-1}$. (b)-(d) Several multiplets of Dy$^{3+}$ at 5 K and (b) at different temperatures. Lines in (b) are labelled by numerals and capital letters indicating initial and final levels of transitions, respectively. Inset compares the $I(T)$ and $n(T)$ dependences for the line 2A (see text).

$$n_i = \frac{\exp(-E_i/kT)}{\sum_{j=0}^{4}\exp(-E_j/kT)}. \quad (1)$$

Here $k$ is the Boltzman constant.

Figure 2 demonstrates typical temperature dependences of spectral lines and their positions and widths. One can clearly see that at $T_N^{Mn}$ the lines experience only a small peculiarity in their positions and a pronounced narrowing at further cooling. No line splitting is observed. Narrowing of the lines correlates with establishment of a magnetic order in the Mn subsystem, accompanied by slowing down fluctuations of the effective magnetic field at Dy sites, which cause a dynamic Zeeman splitting of Dy Kramers doublets and, hence, a line broadening. An absence of line splittings is in agreement with an absence of any induced Dy magnetic moment in the temperature interval $T_{lock} < T < T_N^{Mn}$ [8].

**Table 1** Energies (cm$^{-1}$) of CF levels of Dy$^{3+}$ in paramagnetic DyMnO$_3$. The values in parentheses are from Ref. [14].

| $^6H_{15/2}$ | $^6H_{13/2}$ | | $^6H_{11/2}$ | $^6H_{9/2}+^6F_{11/2}$ | $^6H_{7/2}+^6F_{9/2}$ |
|---|---|---|---|---|---|
| 0 | 3553 | (3551) | 5932 | 7706 | 9060 |
| 107 | 3614 | | 5983 | 7733 | 9108 |
| 167 | 3633 | (3627) | 6030 | 7765 | 9163 |
| 193 (185) | 3702 | (3697) | 6048 | 7808 | 9196 |
| 245 | 3774 | (3776) | 6070 | 7862 | 9225 |
| | 3805 | | 6183 | 7901 | 9281 |
| | 3828 | (3826) | | 7944 | 9350 |
| | | | | 7995 | 9458 |
| | | | | 8044 | |
| | | | | 8093 | |

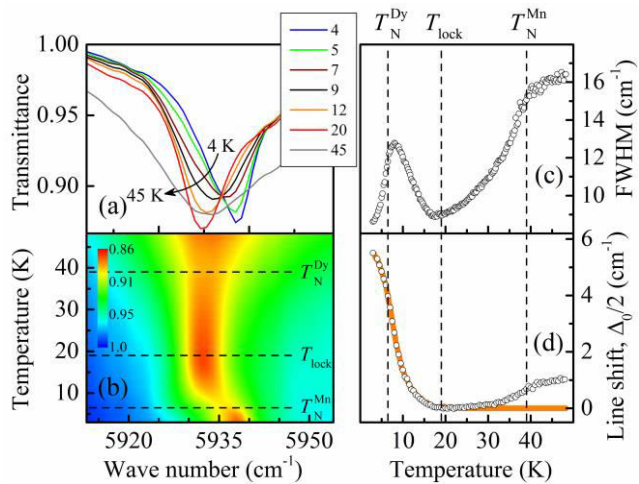

**Figure 2** The temperature evolution of the lowest-frequency Dy$^{3+}$ line in the $^6H_{11/2}$ multiplet presented as (a) transmittance line shapes at several temperatures, (b) intensity map in wave number – temperature coordinates, (c) full width at half maximum (FWHM) and (d) line shift as functions of temperature (symbols). A solid orange line in (d) represents a half of the ground-state splitting, $\Delta_0(T)/2$, see text.

Below $T_{lock}$=18 K, the Dy lines broaden. At temperatures in the vicinity of $T_N^{Dy}$ = 6.5 K, the line width reaches its maximum, a shoulder develops at the low-frequency side of the line, it diminishes in intensity and finally disappears at the lowest temperatures, which results in a decreasing of the line width (see Figs. 2a, 2c). Such changes are typical for all spectral lines registered in the experiment. We interpret them in the following way. The line broadening below $T_{lock}$=18 K comes from a splitting of Dy$^{3+}$ Kra-mers doublets caused by the DM interaction with the *c* component of Mn spins that emerges just below $T_{lock}$. Indeed, at temperatures between $T_N^{Mn}$ and $T_{lock}$, the Mn magnetic moment directed along the *b* axis [8] could induce a collinear magnetic moment of dysprosium via the symmetric Mn-Dy exchange interaction. However, the Dy magnetic moment [8] and the splitting of Dy spectral lines are zero in this interval of temperatures. Below $T_{lock}$, the Dy magnetic moment lying in the *ab* plane (almost along the *b* axis [11]) appears, together with emergence of the *c* component in the Mn magnetic moment. All this indicates that the Dy-Mn exchange is dominated by the antisymmetric DM interaction. On the contrary, a collinear AFM ordering of the Dy subsystem at $T_N^{Dy}$ = 6.5 K is mediated, mainly, by a symmetric Dy-Dy exchange (according to our estimate, the magnetic dipole-dipole interaction is of about 1.5 K).

In general, each spectral line should split into four components in a magnetically ordered state, two of them freezing out at low temperatures (see the scheme of Fig. 3a). Because of broad lines, we observe an averaged picture. Freezing out of the low-frequency components results in an observable line narrowing and shift. We find that all the observed Dy lines behave in the same way. Shifts of different lines differ within ±1 cm$^{-1}$. This value, evidently, includes excited state splittings and shifts due to changes of the crystal field caused by the electric polarization, whereas the main contribution to the shift comes from the ground-state splitting. We also note that while the electric polarization sharply drops for temperatures below $T_{Dy}$=6.5 K, the line shift continues to grow, which gives an additional support to our arguments. Under the assumption that that the measured line shift (Fig. 2d) equals 0.5 $\Delta_0(T)$, we have estimated the Dy ground-state splitting $\Delta_0$ as 11 ± 2 cm$^{-1}$ at 4 K.

It is worth noting that the spectral lines corresponding to the *f-f* electronic transitions of the Dy$^{3+}$ ions are unusually broad in DyMnO$_3$ even in a magnetically ordered state (FWHM ≥ 8 cm$^{-1}$), as compared with other Dy magnetic oxides (~ 3 cm$^{-1}$ in DyFe$_3$(BO$_3$)$_4$ [16]), ~0.2 cm$^{-1}$ in Dy$_2$BaCuO$_5$ [17]). Two reasons could be responsible for that, namely, (i) a large inhomogeneous broadening or, more probably, (ii) the incommensurate magnetic structure of the Mn$^{3+}$ magnetic subsystem, which creates different effective magnetic fields at different Dy sites. The same situation holds for TbMnO$_3$ [18].

Next, we use the $\Delta_0(T)$ dependence, obtained as described above, to estimate the temperature behaviour of Dy$^{3+}$ magnetic moments. In the frame of the ground-doublet model, the relation

$$m_{Dy}(T)/m_{Dy}(0) = (n_1 - n_{1'}) = \tanh(-\Delta_0(T)/2kT) \quad (2)$$

is valid, where $n_1$ and $n_{1'}$ are the populations of the split components of the Dy$^{3+}$ ground doublet. We calculate $m_{Dy}(T)/m_{Dy}(0)$, according to Eq. 2, using the Boltzman distribution (Eq. 1) for the populations of the levels 1 and 1′ separated by the energy interval $\Delta_0(T)$. The resulting curve is shown by a black line in Fig. 3a, where it is compared with the intensities of magnetic peaks found in experiments on x-ray resonant scattering [8] corresponding to Dy magnetic moments in ICM ($T>T_N^{Dy}$) and CM ($T<T_N^{Dy}$) magnetic structures of the Dy subsystem. In optical experiment, the integral splitting is registered, it is not possible to separate contributions of the two magnetic structures. Nevertheless, the calculated curve is qualitatively consistent with the neutron data.

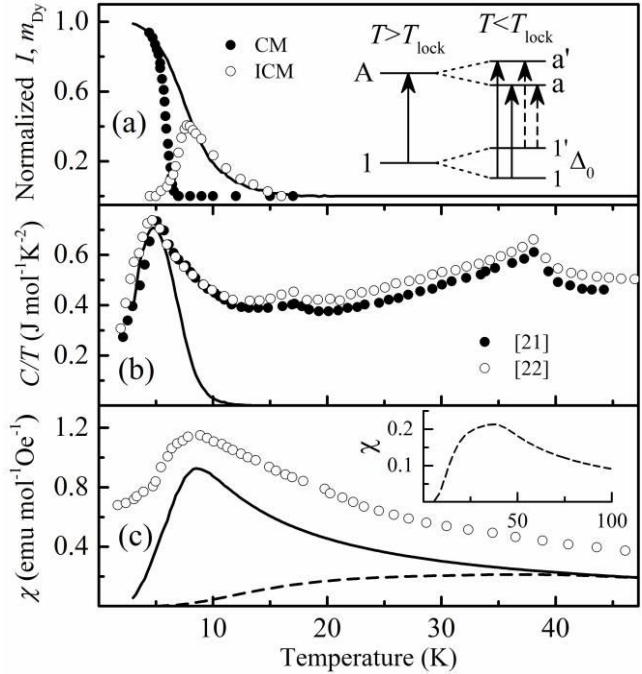

**Figure 3** Dysprosium contribution (solid line) into the (a) magnetic moment, (b) specific heat, and (c) magnetic susceptibility of $DyMnO_3$, calculated from the experimentally found ground-state splitting of $Dy^{3+}$ and compared with (a) x-ray resonant magnetic scattering [8], (b) specific heat [21,22], and (c) magnetic susceptibility (H||b, [22]) measurements. Symbols represent the experimental data. Inset in (a) illustrates Dy line splitting in a magnetically ordered state. Inset and the dashed line in (c) display a hypothetical Dy contribution in the case of $\Delta_0(0) = 30$ cm$^{-1}$ [14].

To additionally check the $\Delta_0(T)$ dependence estimated from the optical experiment, we have calculated the dysprosium contributions to the magnetic susceptibility $\chi_{Dy}(T)$ and specific heat $C_{Dy}(T)$, according to the following formulae [19,20]:

$$\chi_b^{Dy}(T) = N_A \frac{m_b^{Dy}(0)^2}{kT} \frac{1}{\cosh^2(\Delta_0(T)/2kT)}, \quad (3)$$

$$C_{Dy}(T) = R \left(\frac{\Delta_0(T)}{2kT}\right)^2 \frac{1}{\cosh^2(\Delta_0(T)/2kT)} \quad (4)$$

Here $m_b^{Dy}(0)$ is the b-component of the dysprosium magnetic moment, $N_A$ and $R$ are the Avogadro number and the universal gas constant, respectively. The dependences calculated with the aid of Eqs. 3 and 4 are compared with the experimental data from literature in Figs. 3c and 3b, respectively. The contributions of dysprosium describe reasonably well low-temperature anomalies in $\chi(T)$ and $C(T)$ reported in the experiments on $DyMnO_3$. We note that $\Delta_0(0) = 30$ cm$^{-1}$ [14] fails to explain the measured $\chi(T)$ dependence (Fig. 3c).

Interestingly, the calculated Schottky-type dysprosium contribution $C_{Dy}(T)$ describes almost the whole experimental peak at about 5 K corresponding to $T_N^{Dy}$ [21,22]. A similar Schottky-type peculiarity was observed near $T_N^{Gd}$ in the $C(T)$ dependence measured for $GdMnO_3$ [23]. However, in that case, a very narrow sharp spike at $T_N^{Gd}$ was superimposed onto the Schottky-type anomaly [23]. Possibly, in the case of $DyMnO_3$, a sharp peak also existed but could not be registered because of, e.g., a rather poor quality of the sample. The latter follows from not so sharp shapes of the other two peculiarities in $C(T)$, Fig. 3b, as compared to respective features measured for $GdMnO_3$ of a perfect quality [23].

**4 Conclusion** To summarize, we have undertaken a thorough high-resolution spectroscopic investigation of $DyMnO_3$ single crystals in a wide range of wave numbers (2000 - 10000 cm$^{-1}$) and temperatures (5 – 300 K). We show that a broad absorption band corresponding to a d-d optical transition of $Mn^{3+}$ and narrow lines due to f-f transitions of $Dy^{3+}$ are sensitive to magnetic ordering phase transitions in both manganese and dysprosium subsystems, thus manifesting a mutual interaction of these subsystems.

We have determined energies of 31 out of 33 crystal-field levels of the excited Dy multiplets $^6H_{13/2}$, $^6H_{11/2}$, $^6H_{9/2}+^6F_{11/2}$, and $^6H_{7/2}+^6F_{9/2}$ and 5 out of 8 levels of the ground $^6H_{15/2}$ multiplet of a paramagnetic $DyMnO_3$, which forms a basis for future reliable crystal-field calculations. As the first excited CF level lies at 107 cm$^{-1}$, the low-temperature properties of the Dy subsystem are governed by the ground Kramers doublet split in the magnetically ordered states.

We show that the Dy-Mn exchange is dominated by the antisymmetric DM interaction. On the contrary, a collinear AFM ordering of the Dy subsystem at $T_N^{Dy} = 6.5$ K is mediated, mainly, by a symmetric Dy-Dy exchange

A study of the line shapes, widths, and shifts enabled us to estimate the $Dy^{3+}$ ground Kramers doublet splitting as 11±2 cm$^{-1}$ at 4 K and to find the $\Delta_0(T)$ dependence. We used $\Delta_0(T)$ to calculate the Dy magnetic moment $m_{Dy}(T)$ and the Dy contribution into the magnetic susceptibility and specific heat of $DyMnO_3$.

We note that the electronic excitation of Dy corresponding to the optical transition between the split components of the Dy ground-state Kramers doublet falls into the energy region of magnons in $DyMnO_3$ [9]. It would be of interest to study a possible interaction between these two kinds of excitations in multiferroic $DyMnO_3$ by terahertz spectroscopy or neutron scattering experiments.

**Acknowledgements** Financial support of the Russian Academy of Sciences under the Programs for Basic Research, of the Russian Foundation for Basic Research (grant No. 14-02-90467), and of State Contract (project No. 872) is acknoweledged.

**References.**

[1] S.-W. Cheong and M. Mostovoy, Nat. Mater. **6**, 13 (2007).
[2] M. Kenzelmann, A. B. Harris, S. Jonas, C. Broholm, J. Schefer, S. B. Kim, C. L. Zhang, S.-W. Cheong, O. P. Vajk, and J. W. Lynn, Phys. Rev. Lett.**95**, 087206 (2005).
[3] T. Arima, A. Tokunaga, T. Goto, H. Kimura, Y. Noda, and Y. Tokura, Phys. Rev. Lett. **96**, 097202 (2006).
[4] I. A. Sergienko and E. Dagotto, Phys. Rev. B **73**, 094434 (2006).
[5] J. A. Alonso, M. J. Martínez-Lope, M. T. Casais, and M. T. Fernández-Díaz, Inorg. Chem. **39**, 917 (2000).
[6] T. Kimura, G. Lawes, T. Goto, Y. Tokura, and A. P. Ramirez, Phys. Rev. B **71**, 224425 (2005).
[7] R. Feyerherm, E. Dudzik, N. Aliouane, and D. N. Argyriou, Phys. Rev. B **73**, 180401(R) (2006).
[8] O. Prokhnenko, R. Feyerherm, E. Dudzik, S. Landsgesell, N. Aliouane, L. C. Chapon, and D. N. Argyriou, Phys. Rev. Lett. **98**, 057206 (2007).
[9] T. Finger, K. Binder, Y. Sidis, A. Maljuk, D. N. Argyriou, and M. Braden, Phys. Rev. B **90**, 224418 (2014).
[10] M. Schiebl, A. Shuvaev, A. Pimenov, G. E. Johnstone, V. Dziom, A. A. Mukhin, V. Y. Ivanov, and A. Pimenov, Phys. Rev. B **91**, 224205 (2015).
[11] R. Feyerherm, E. Dudzik, A. U. B. Wolter, S. Valencia, O. Prokhnenko, A. Maljuk, S. Landsgesell, N. Aliouane, L. Bouchenoire, S. Brown, D. N. Argyriou, O. Prokhnenko, and D. N. Argyriou, Phys. Rev. B **79**, 134426 (2009).
[12] N. Zhang, S. Dong, Z. Fu, Z. Yan, F. Chang, and J. Liu, Sci. Rep. **4**, 6506 (2014).
[13] N. Kida, Y. Ikebe, Y. Takahashi, J. P. He, Y. Kaneko, Y. Yamasaki, R. Shimano, T. Arima, N. Nagaosa, and Y. Tokura, Phys. Rev. B **78**, 104414 (2008).
[14] S. Jandl, S. Mansouri, A. A. Mukhin, V. Yu Ivanov, A. Balbashov, M. M. Gospodino, V. Nekvasil, and M. Orlita, J. Magn. Magn. Mater. **323**, 1104 (2011).
[15] A.M.Balbashov and S.K.Egorov, J.Cryst. Growth, **52**, 498 (1981)
[16] M. N. Popova, unpublished results.
[17] M. Baran, H. Szymczak, S. A. Klimin, M. N. Popova, R. Z. Levitin, and B. V. Mill', J. Exp. Theor. Phys. **84**, 175 (1997).
[18] M.A. Kashchenko, S.A. Klimin, Optics and Spectroscopy, **116**, 836 (2014).
[19] E.A. Popova, S.A. Klimin, M.N.Popova, R. Klingeler, N.Tristan, B.Büchner, A.N.Vasiliev, J. Magn. Magn. Materials **331**, 133 (2013).
[20] S.A. Klimin, A.S. Galkin, M.N. Popova, J. Alloys Comp. **625**, 193 (2015).
[21] T. Zou, Z. Dun, H. Cao, M. Zhu, D. Coulter, H. Zhou, and X. Ke, Appl. Phys. Lett. **105**, 052906 (2014).
[22] A. Midya, S. N. Das, P. Mandal, S. Pandya, and V. Ganesan, Phys. Rev. B **84**, 235127 (2011).
[23] A. A. Wagh, K. G. Suresh, P. S. Anil Kumar, and S. Elizabeth, J. Phys. D. Appl. Phys. **48**, 135001 (2015).